\documentclass[11pt,twocolumn,twoside]{IEEEtran}
\usepackage{amsmath}
\usepackage[margin=0.75in,headheight=0.45in]{geometry}
\usepackage[pdftex]{epsfig}
\usepackage{amsfonts}
\usepackage{amssymb}
\usepackage{fancyhdr}
\include{graphicsx}

\usepackage{multirow}
\usepackage{siunitx}

\newcommand{\norm}[1]{\lVert#1\rVert}

\begin{document}
\title{\vspace{0.2in}\sc Dimensionality-Reduction of Climate Data using Deep Autoencoders}
\author{Juan A. Saenz$^{1}$\thanks{Corresponding author: J.A. Saenz, juan.saenz@lanl.gov; N. Lubbers, nlubbers@bu.edu; N.M. Urban, nurban@lanl.gov.} \thanks{$^1$Los Alamos National Laboratory, $^2$Boston University.}, Nicholas Lubbers$^{1,2}$, Nathan M. Urban$^{1}$}

\maketitle
\thispagestyle{fancy}
\begin{abstract}
We explore the use of deep neural networks for nonlinear dimensionality reduction in climate applications. We train convolutional autoencoders (CAEs) to encode two temperature field datasets from pre-industrial control runs in the CMIP5 first ensemble, obtained with the CCSM4 model and the IPSL-CM5A-LR model, respectively.
With the later dataset, consisting of 36500 96$\times$96 surface temperature fields, the CAE out-performs PCA in terms of mean squared error of the reconstruction from a 40 dimensional encoding.
Moreover, the noise in the filters of the convolutional layers in the autoencoders suggests that the CAE can be trained to produce better results. Our results indicate that convolutional autoencoders may provide an effective platform for the construction of surrogate climate models.
\end{abstract}

\section{Introduction}
Uncertainty quantification of the response of the Earth system to greenhouse-gas emission scenarios is important for evaluating the impacts of climate change on infrastructure, agriculture, and the environment, among other areas.
However, simulations using global, coupled earth system models are computationally expensive, making it impossible to produce large ensembles needed for statistical uncertainty quantification.
To overcome this, surrogate models--simplified models that emulate more complex climate models--are built and trained \cite{little_etal_2013, little_etal_2013b}.
The computational cost of running these climate model emulators is much lower than their full complexity counterparts. 
As a result, large ensembles of simulations (order tens-hundreds of thousands) can be produced and used to carry out uncertainty quantification.

Emulators can be devised to dynamically evolve the state of the climate on a dimensionally reduced manifold \cite{ross_2009}.
An important requirement of such emulators is that the state in physical dimensions can be recovered.
Linear dimensionality reduction via principal component analysis (PCA) is well-known in the climate science community, however, nonlinear methods have not been fully explored.
Ross \cite{ross_2009} investigated nonlinear dimensionality reduction methods for a different climate aplication: identifying low-dimensional nonlinear dynamics in El Nino variability. Errors from reconstructions using nonlinear methods were not significantly better than using linear PCA. Methods include nonlinear PCA (autoencoders), Isomap, and Hessian locally linear embedding. %Consider citations!

Here, motivated by the success of deep autoencoders for dimensionality reduction~\cite{hinton_salakhutdinov_2006} and convolutional neural networks for image processing~\cite{alexnet,vggnet}, we present work in progress using convolutional autoencoders~\cite{convautoencoder} to reduce the dimensionality of data from climate models. In section~\ref{s:methods} we describe our methods, and in section~\ref{s:results} we present results on two pre-industrial climate model simulation datasets: the CCSM4-T31 temperature at the surface dataset, and the IPSL-CM5A-LR temperature at the surface data. We end with a brief discussion in section \ref{s:discussion}.

\section{Methods}
\label{s:methods}
\subsection{Principal Components Analysis}
\label{s:pca} 
Consider a dataset of dimensionality $M$ with $N$ datapoints collected into a $N \times M$ data matrix $\mathbf{X}$. PCA constructs a rank $m$ reduced matrix $ \hat{\mathbf{X}}_{pca}$ by projecting $\tilde{ \mathbf{X} }$ (obtained by centering and normalizing $\mathbf{X}$ using the global mean and standard deviation) onto the first $m$ principal components which maximizes the covariance of the data, thus minimizing the mean reconstruction error
\begin{equation}
\label{eq:mse}
MSE = \frac{1}{NM} \sum_n{(\mathbf{x}_n-\mathbf{\hat{x}}_n)^2} =\frac{1}{NM} {\norm{\mathbf{X}-\hat{\mathbf{X}}}_2^2}.
\end{equation}
This can be obtained by singular value decomposition of $\mathbf{X}$ = $\mathbf{U}\mathbf{\Sigma}\mathbf{W}^T$ with $\mathbf{\Sigma}$ the diagonal matrix of singular values. The data covariance is given by $\mathbf{X}^T \mathbf{X}$ = $\mathbf{W} \mathbf{\Sigma}^2 \mathbf{W}^T$, and so the first $m$ principal component vectors are the columns of $\mathbf{W}$ associated with the $m$ largest singular values.

\subsection{Convolutional Autoencoder}
\label{s:convauto}

The autoencoder provides an alternative method for dimensionality reduction of $\mathbf{X}$. 
Data is fed though a series of neural network layers, i.e. an affine transform followed by an elementwise nonlinearity $f$, to create activations $\mathbf{X}_l$ at layer $l$:
\begin{equation}
\mathbf{X}_l = f(\mathbf{W}_l \mathbf{X}_{l-1}+\mathbf{b}_l).
\end{equation}
The trainable parameters of the network are the weights $\mathbf{W}_l$ and biases $\mathbf{b}_l$ of each layer. The waist layer of the autoencoder is constrained to a number of neurons $m$, so that the activations at that layer can be used as an $m$ length code for each image. This is followed by decoding layers, and in the final layer of the autoencoder, $\mathbf{\hat{X}}$ is constructed to have the same dimensionality as the input $\mathbf{X}$. The parameters of the network are then trained to minimize the same reconstruction error $MSE$ (Eq.~\ref{eq:mse}) as PCA.

The convolutional autoencoder (CAE) is an extension to the autoencoder which faciltates the analysis of data on regular grids, such as images. Here, each data point can be indexed by two pixel positions. In the convolutional and deconvolutional layers, the weights consist of small image filter kernels, and the product of the weights and the data consist of spatial convolutions. Denoting the feature $k$ of the pixel indexed by $i$ and $j$ by $\mathbf{x}_{i,j}^k$, the preactivations $\mathbf{W}\mathbf{x}$ are computed using
\begin{equation}
(\mathbf{W} \mathbf{x})_{i,j}^l = \sum_{k,l,m}{\mathbf{W}^{lk}_{{i-l},{j-m}} \mathbf{x}^k_{i,j}} \,. 
\end{equation}
Convolutional layers operate as a set of local image filters with the capacity to extract patterns that increase in complexity with depth \cite{convpatterns}. They have the benefit of greatly reducing the number of learnable parameters per layer, and faciliate training by seeking out features which encode the structure of local image patches \cite{lecun98}.

Finally, CAEs have pooling layers that coarse-grain the image plane after each convolution. We use $2\times2$ max pooling, with unpooling layers which use piece-wise constant $2\times2$ upsampling.
We use two regularization strategies to improve training regularity and smoothness of learned filters. 
In the first, we use image flipping and weight decay with strength $\beta$. In the second, we use noise injection (denoising autoencoder), applying pixel-wise Gaussian noise with mean $0$ and standard deviation $\gamma$ to the images. 

Autoencoders presented here are trained using stochastic gradient descent for 1000 epochs with a learning rate of 0.01, which is updated using a Nesterov scheme with a momentum of 0.975, and a batch size of 128. 
Convolutional layers use linear activation functions (we have not been able to successfully train non-linear activations). 
Fully connected layers use rectified linear (ReLU) activation functions. 
Decoding layers have the reverse structure of the encoding layers, but we do not tie the weights between the encoding and decoding layers. Convolutional boundary conditions use valid convolutions (no padding), with decoding convolutional layer dimensions computed to produce the correct size output. As preprocessing, the temperature fields are normalized to mean $0$ and standard deviation $1$ using the global mean and standard deviation.

% map between notation used in this paper and the notation used in training.
% A1 = A12 
% A2 = A15 
% B1 = A25 
% B2 = A26.01 
% B3 = A26.02 
\begin{table}[]
\caption{Architectures of Convoluational Autoencoders}
\vspace{-10pt}
\label{t:architectures}
\begin{center}
\begin{tabular}{ c | c | c  c}
label & encoding architecture & $\beta$ & $\gamma$ \\
\hline
A1 & CL7,32-PL-FC40 & 0.00025 & N/A \\
A2 & CL5,32-PL-CL5,64-PL-FC40 & 0.00025 & N/A \\
B1 & CL5,32-PL-CL5,64-PL-FC40 & N/A & N/A \\
B2 & CL5,32-PL-CL5,64-PL-FC40 & N/A & 0.1\\
B3 & CL5,32-PL-CL5,64-PL-FC40 & N/A & 0.5\\
\end{tabular}
\vspace{5pt}

CL$n$,$m$ is a convolutional layer with an $n\times n$ receptive field and $m$ features, PL is a 2$\times$2 pooling layer, and FC$m$ is a fully connected layer with $m$ neurons.
\end{center}
\vspace{-20pt}
\end{table}

\section{Results}
\label{s:results}

\subsection{CCSM4}
\label{s:CCSM4_T31}

%project=PMIP3, model=NCAR, National Center for Atmospheric Research, USA, experiment=Pre-Industrial Control, time_frequency=monClim, cmor_table=Lclim, modeling realm=land, ensemble=r1i1p1, version=20140428 
%r1i1p1: first ensemble member.

We train encoders with data from the Community Climate System Model - version 4 (CCSM4)~\cite{gent_etal_2011}.
We employ 150 years of surface temperature ($T_s$) monthly climatology data from the first ensemble, pre-industrial control run of the 5th version of the Climate Model Intercomparison Project (CMIP5).
The data was regridded to a 3.75$\times$3.75 degree grid (T31 grid).
The resulting dataset has 1800 samples of 48$\times$96 $T_s$ fields.

%\begin{figure}
%\begin{center}
%\includegraphics[trim={0 28pt 0 0}, clip, width=0.8\hsize,angle=0]{X_pred_test_T31_l12.pdf} %\\
%%\vspace{2pt}
%%\includegraphics[trim={0 0 0 150pt}, clip, width=0.95\hsize,angle=0]{X_pred_test_T31_l15.pdf}
%\end{center}
%\caption{Comparison of $T_s$ [\si{\degree}K] from the original dataset (top) and the reconstructed with A12 (bottom), sample 937 from the CCSM4-T31 dataset.}
%\label{f:T31}
%\end{figure}

\begin{figure}
\begin{center}
\includegraphics[width=0.3\hsize,angle=0]{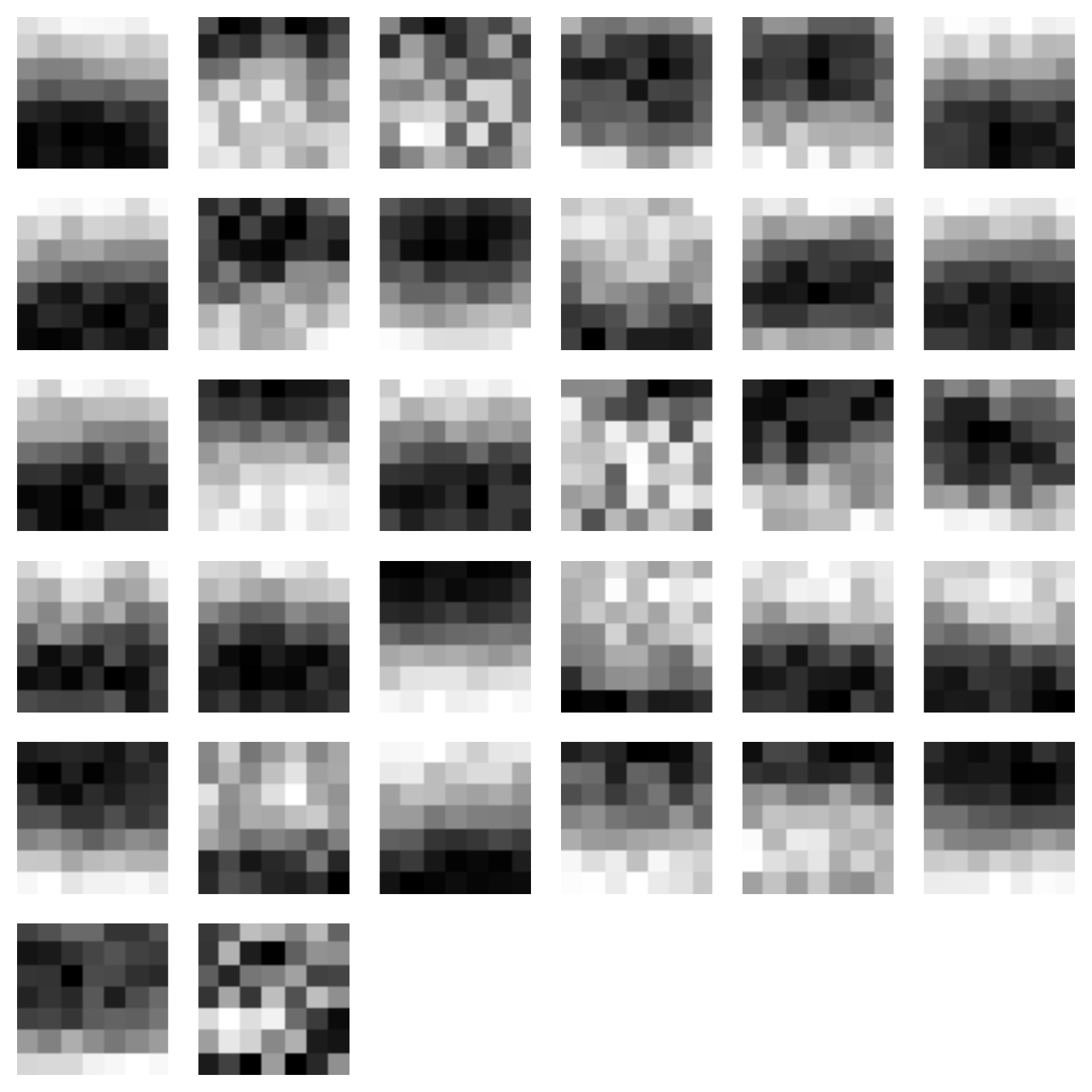}
\hspace{0.05\hsize}
\includegraphics[width=0.3\hsize,angle=0]{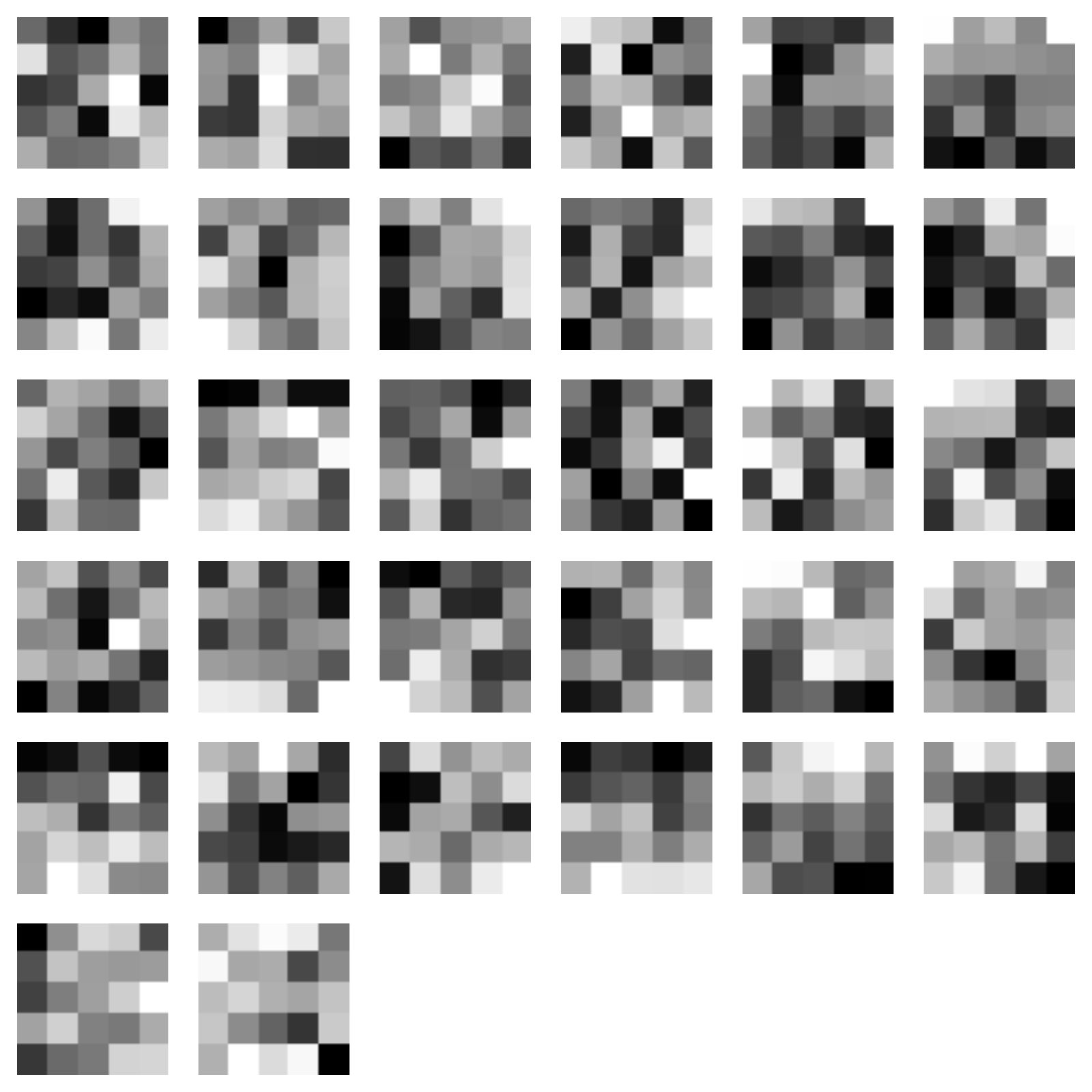} \\
\vspace{0.02\hsize}
\includegraphics[width=0.3\hsize,angle=0]{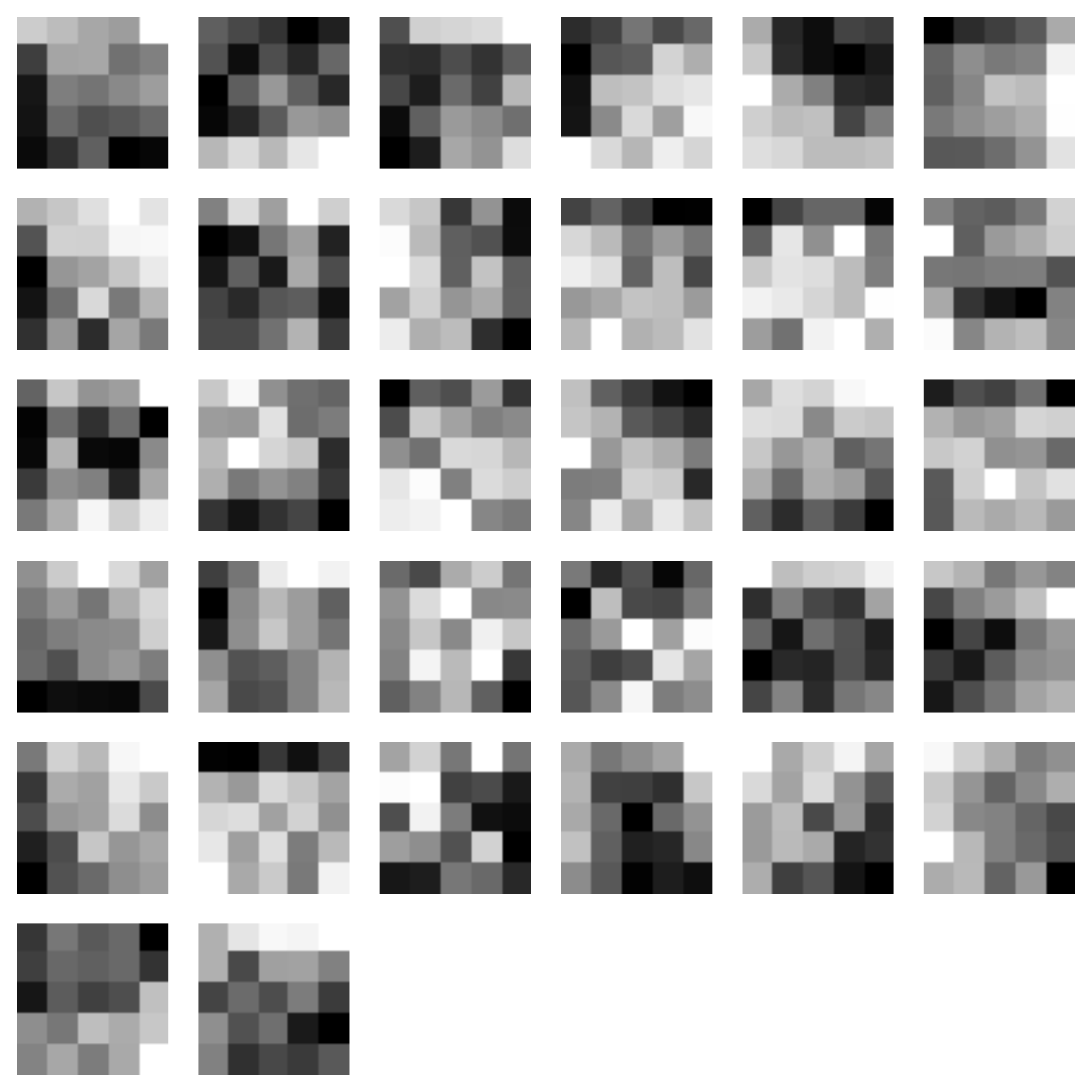}
\hspace{0.05\hsize}
\includegraphics[width=0.3\hsize,angle=0]{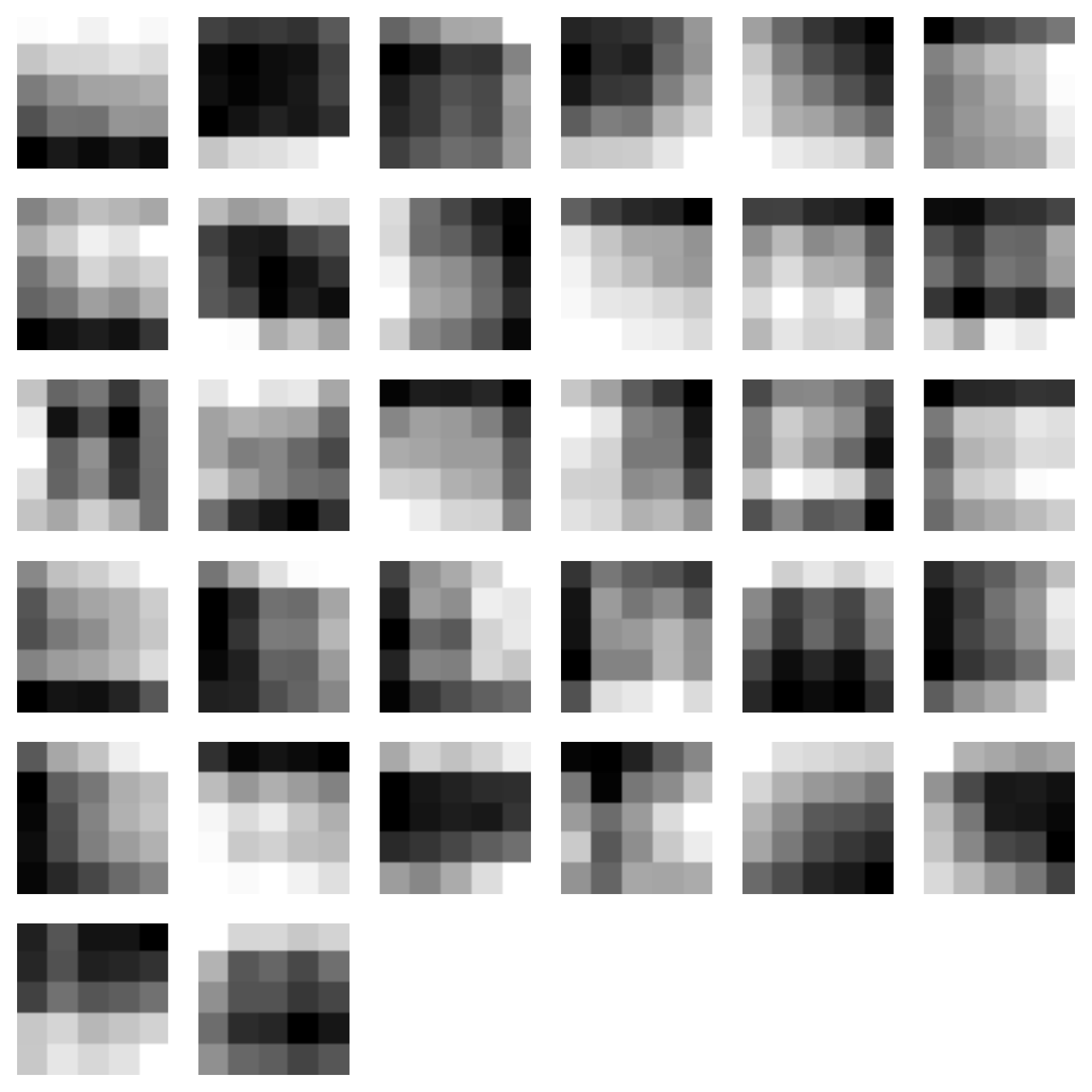} \\
\end{center}
\caption{Weights from the first convolutional layer ($\mathbf{W}^1$) in the trained architectures A1 (top left) and B1 (top right), B2 (bottom left) and B3 (bottom right). Brightness scale is arbitrary, only the patterns visible in the plots are meaningful here.}
\label{f:weights}
\end{figure}

The trained architectures (A1 and A2, table~\ref{t:architectures}) do a good job at recovering the global structure of the temperature field $T_s$. 
There are differences in local features in some regions (not shown).
In table \ref{t:mse} we compare the $MSE$ (Eq.~\ref{eq:mse}) of reconstructions using CAEs A1 and A2 and PCA. 
Architecture A1 performs better than A2, but PCA performed better than both autoencoders. 
The weights of the first convolutional layer indicated that this type of regularization was not effective (Fig.~\ref{f:weights}, upper left). 
There is some repetitive structure between weights, and they are noisy, indicating that the neural networks are not well trained. 
One possible reason for this is that there are not enough samples ($N$=1800). 
To investigate this, we now use a dataset with a much larger number of samples.

%l12 = A1
%# Neural Network with 3019433 learnable parameters
%name          size         total
%------------  ---------  ------- 
%input         1x48x96       4608   
%conv          32x42x90    120960   
%pool          32x21x45     30240   
%flatten       30240        30240   
%encode_layer  40              40  
%hidden        44064        44064  
%unflatten     32x27x51     44064 
%unpool        32x54x102   176256
%deconv        1x48x96       4608   
%output_layer  4608          4608   

%l15 = A2
%# Neural Network with 1650729 learnable parameters
%name          size         total  
%------------  ---------  -------  
%input         1x48x96       4608   
%conv1         32x44x92    129536  
%pool1         32x22x46     32384   
%conv2         64x18x42     48384    
%pool2         64x9x21      12096    
%flatten       12096        12096  
%encode_layer  40              40  
%hidden        25920        25920  
%unflatten     64x15x27     25920  
%unpool3       64x30x54    103680  
%deconv3       32x26x50     41600  
%unpool4       32x52x100   166400  
%deconv4       1x48x96       4608  
%output_layer  4608          4608  

\subsection{IPSL-CM5A-LR}
\label{s:IPSL}

Here, we use near-surface air temperature $T_{as}$ from the low resolution pre-industrial control run produced by the IPSL-CM5A-LR model, part of the CMIP5 first ensemble \cite{dufresne_etal_2013}.
The data is on a 1.9$\times$3.75 degree grid, 96$\times$96, and we use daily output for 100 years between 1800 and 1900 ($N$=36500 samples), although 600 years of data are available (219000 samples).
As a result, the dataset has 36500 samples of 96$\times$96 $T_{as}$ fields, resulting in $\mathbf{X}$ with shape $(N=36500) \times (M=9216)$.

We implement an autoencoder architecutre, B1, which is similar to A2, but without any type of regularization (table \ref{t:architectures}). The MSE using B1, shown in table \ref{t:mse}, is smaller than the error obtained with PCA. However, the weights are still noisy (Figure \ref{f:weights}, upper right).
To remedy this, we explored architectures B2 and B3 (table \ref{t:architectures}) which are regularized using injected noise. The weights become smoother with more noise (Fig~\ref{f:weights}, lower left and right), but the errors are larger, as shown in table \ref{t:mse}.

\begin{figure}
\begin{center}
\includegraphics[trim={0 25pt 0 0}, clip, width=0.8\hsize,angle=0]{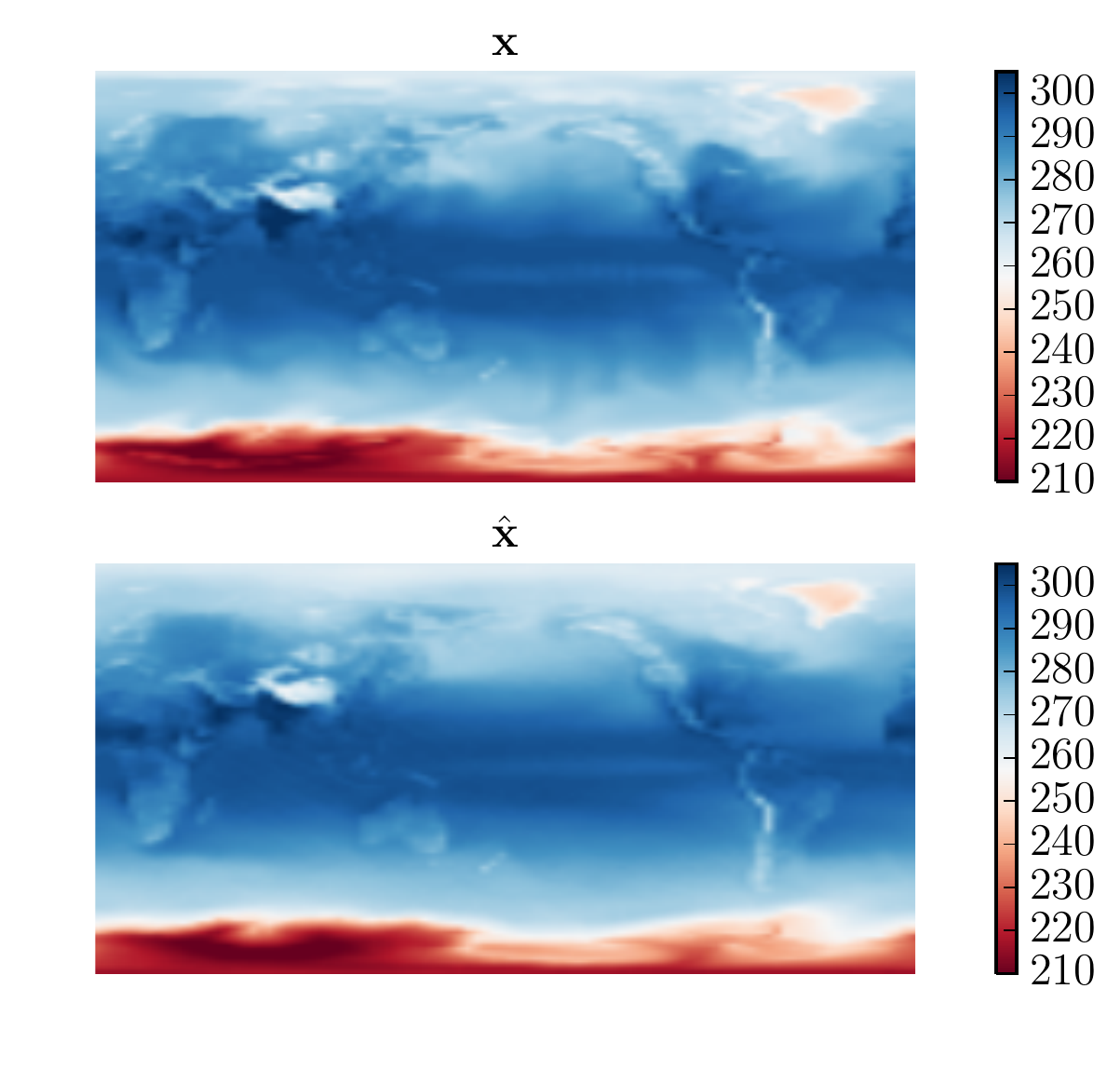}\\
\vspace{2pt}
\includegraphics[trim={0 25pt 0 155pt}, clip, width=0.8\hsize,angle=0]{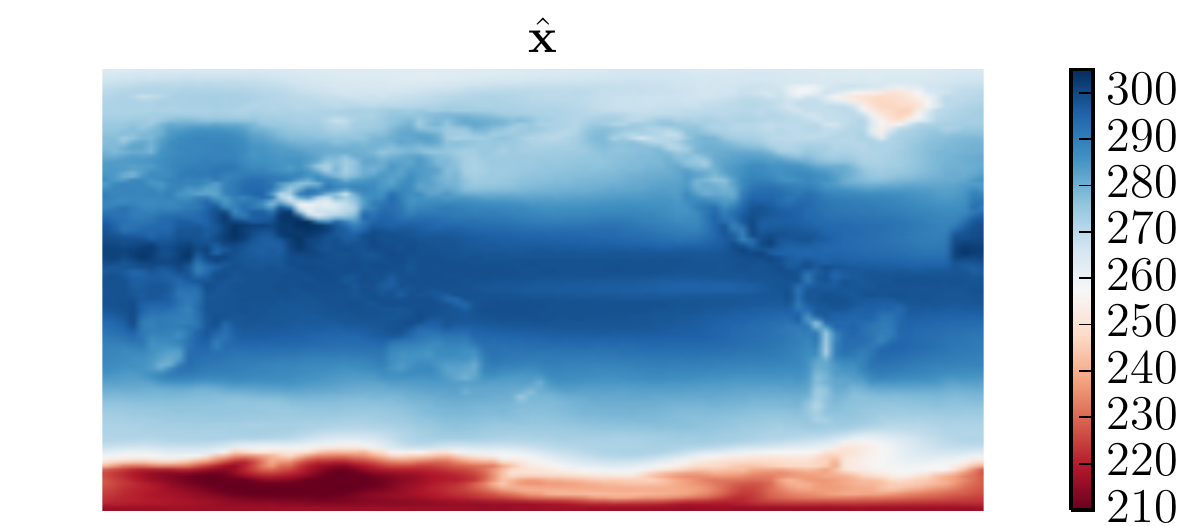}
\end{center}
\caption{Comparison of $T_{as}$ [\si{\degree}K] from the original dataset (top) and the reconstruction using architectures B1 (middle) and B2 (bottom), sample 28618 from the IPSL-CM5A-LR dataset.}
\label{f:IPSL}
\end{figure}

In figure \ref{f:IPSL} we show the reconstructed temperature fields, which are very similar to the temperature in the original dataset.
Large scale features of the global temperature patterns are preserved.
Smaller scale features in regions such as over the Antarctic peninsula, the North Atlantic and the South Pacific are filtered out.
Some small scale features above high elevation topography, such as the Andes and the Himalayas,  appear to be well preserved.

% probably better to show rms errors?
%\begin{table}
%\begin{center}
%  \caption{} 
%  \label{t:T31}
%  \vspace{-10pt}
%  Mean squared error for architectures trained on the CCSM4-T31 dataset.
%  \\
%  \vspace{5pt}
%  \begin{tabular}{ c c c c }
%    \hline
%    PCA 		& A12		& A15 	&		\\ \hline 
%    0.8022	& 1.5495 		& 2.1608	&		\\ \hline \hline
%    PCA		& A25		& A26.01	& A26.02	\\ \hline
%    4.9014	& 4.2415		& 4.3806	&     		\\ \hline \hline
%  \end{tabular}
%\end{center}
%\end{table}

\begin{table}[]
\caption{Mean squared error of reconstructions [\si{\degree}K$^2$].}
\vspace{-10pt}
\label{t:mse}
\begin{center}
\begin{tabular}{ c | cccc }
\multirow{2}{*}{CCSM4-T31} & PCA & A1	 & A2 & \\ 
 & 0.8022 & 1.5495 &  2.1608 & \\ \hline
\multirow{2}{*}{IPSL-CM5A-LR} & PCA & B1 & B2 & B3 \\
 & 4.9014 & 4.2415 &  4.3806 & 4.5243 
\end{tabular}
\end{center}
\vspace{-20pt}
\end{table}

%PCA: mse = 4.901439
%A25:  mse = 4.241502996774634
%A26.01: mse = 4.3806465612062224

\section{Discussion}
\label{s:discussion}

The results of this work in progress indicate that there is potential to devise deep autoencoders for dimensionality reduction of climate data. Noise in the filters of the trained networks indicate that finding effective representations is dependent on regularization and the availability of data.

Future developments will focus on testing networks with data not used for training/validating, using larger datasets, using nonlinear convolutional activation functions, implementing other regularization methods (e.g. dropout), and using deeper networks, with the aim of improving the reconstruction of small scale features. Future analysis will include the investigation of patterns extracted by the convolutions.

\section*{Acknowledgments}
This work was supported by the Office of Science (BER), U. S. Department of Energy.
Autoencoders are built using Python libraries Theano~\cite{theano_2016}, Lasagne~\cite{sander_dieleman_2015_27878}, and nolearn~\cite{nolearn}.
We thank A. Jonko for providing the CCSM4-T31 data.
We acknowledge the WCRP-WGCM, which is responsible for CMIP. 
We thank the DOE and NCAR for developing the CCSM4, producing and making available their model output.

\bibliographystyle{ieeetr}
%\bibliography{master_bibliography.bib}

\end{document}